# On the Challenges of Collaborative Data Processing

**Sylvie Noël**
*Communications Research Centre, Canada*
**Daniel Lemire**
*UQAM, Canada*

**ABSTRACT**

The last 30 years have seen the creation of a variety of electronic collaboration tools for science and business. Some of the best-known collaboration tools support text editing (e.g., wikis). Wikipedia's success shows that large-scale collaboration can produce highly valuable content. Meanwhile much structured data is being collected and made publicly available. We have never had access to more powerful databases and statistical packages. Is large-scale collaborative data analysis now possible? Using a quantitative analysis of Web 2.0 data visualization sites, we find evidence that at least moderate open collaboration occurs. We then explore some of the limiting factors of collaboration over data.

**KEYWORDS**

Web-Based Applications, Human-Machine Systems, Data Warehousing, Collaborative Research, Social Web, Computer-Supported Collaborative Work

**INTRODUCTION**

Electronic collaboration tools are widespread. Many of these tools are aimed at supporting either group meetings (brainstorming tools, shared whiteboards, videoconferencing tools) or collaborative writing (wikis). These tools have been studied extensively (Pedersen et al., 1993; Okada et al., 1994; Adler et al., 2006). However, although more and more data is being collected, indexed and made available to all, collaborative data processing has received little attention until recently (Viégas et al., 2007, 2008).

Data analysis is a complex but structured task requiring specialized tools such as spreadsheets or statistical packages, some basic knowledge of statistics and information technology, and the domain knowledge to interpret the results. As opposed to text, scientific or business data is often organized in rigid structures (e.g., tables, lists, networks) and it may be more difficult to interpret without appropriate visualization tools. Regardless of these difficulties, people are interested in viewing and understanding this data. Already people have access to and are familiar with financial and meteorological data, which appear regularly on television, in newspapers and on popular news sites. People are also willing to explore other types of data. For example, a website presenting statistics about baby names proved very popular (Wattenberg, 2005). Businesses of all sizes, governments, and academics analyze data for many purposes: financial planning, sales and marketing, stocks analysis, scientific research, and so on.

In companies, work-related data is called business information. The term "Business Intelligence" (BI) refers to the techniques used to improve decisions by collecting and aggregating business information. BI systems typically use a data warehouse: a large collection of historical and current data on business operations. End-user BI tools include static reports, spreadsheets linked to data repositories and interactive web applications. There is a growing business intelligence industry: the BI market grew by 10% in 2007 alone (Gartner Inc., 2007). One example of a collaborative BI business is Salesforce.com, a SaaS (software as a service) company which helps its customers share various types of business information (Dignan, 2007). Salesforce.com charges a monthly fee to customers to be able to share sales information among themselves.

While companies tend to keep their internal data private to keep an advantage over their competitors, governments and funding agencies increasingly require that scientific data repositories be accessible to all. For example, the Canadian Institutes of Health Research have a policy on Access to Research Outputs which requires grant recipients to deposit data into public databases (Canadian Institutes of Health Research, 2007). Several United Kingdom funding agencies have similar policies, including the Biotechnology and Biological Sciences Research Council, the Economic and Social Research Council, and the Engineering and Physical Sciences Research Council. In 1999, the American Congress passed circular A-110, which extended the Freedom of Information Act (FOIA) to all data produced under a funding award. China plans to make 70% of all scientific data publicly available by 2020 (Niu, 2006). There are a growing number of agencies with Open Access policies, including the U. S. National Institutes of Health, France's Institut national de la santé et de la recherche médicale, Italy's Instituto Superiore di Sanita, Australia's National Health and Medical Research Council, and so on. Some examples of open online scientific databases include the Generic Model Organism Database (Stein et al., 2002), the UK Data Archive for social science data, the Finnish Social Science Data Archive, and Harvard-MIT Data Center. More general open source projects for scientists are also appearing on the web. Examples include OpenWetWare.org (Butler, 2005), Science Commons (Wilbanks & Boyle, 2006), and myExperiment.org. Access to the results of scientific projects has become easier thanks to the proliferation of open access journals; the Directory of Open Access Journals (Lund University Libraries, 2003) lists over 3,000 such journals.

Open database projects also exist outside of the scientific domain, such as Swivel (Swivel Inc., 2007), Freebase (http://www.freebase.com), Numbrary (http://www.numbrary.com), and IBM Many Eyes (IBM, 2007). Amazon makes available large datasets from its web service platform (http://aws.amazon.com/publicdatasets/), including the Human Genome, various US census databases, and various labor statistics. Even the intelligence community, previously focused on secrecy, has been called to focus on information sharing (Jones, 2007). Analysis of the American intelligence efforts to prevent 9/11 has revealed that the lack of information sharing between government agencies left many of them surprised by the attack. There is a call to move from a need-to-know approach to a need-to-share one (Findley & Inge, 2005).

In spite of all this online data, we are not aware of any large-scale collaborative data analysis initiative comparable to those in the fields of software design (open-source software initiatives such as Linux) or documentation (Wikipedia). There might be vast collaborative data-analysis projects, but if there are, they apparently happen behind closed doors or have low visibility.

What is limiting large-scale collaborative data analysis, if anything? Could it be caused by limited accessibility to the data among group members? Is it a lack of collaborative tools or motivation? Or might it be the complexity of the task itself? Is it utopian to think that, one day, experts will collaboratively analyze these data repositories? What conditions must be met, both organizationally and technically, so that collaborative data processing can be fruitful? This chapter presents some of the existing tools that can be used to support collaborative data analysis, explores some of the reasons why people may be refraining from this type of collaboration, and speculates on what we, as researchers and designers, can do to help support this kind of collaboration.

This chapter is divided into the following sections:

- We begin by establishing that the possible lack of collaborative data analysis is not due to the absence of specialized tools. We present and discuss some of these tools, including database tools that store data, and data manipulation tools such as spreadsheets, statistics, and visualization packages. We discuss the recent availability of high quality web-based applications (Descy, 2007), which should make collaborative data analysis more attractive. We also review collaboratories: collaboration tools aimed specifically at scientists.
- We then present the results from an experimental evaluation in which we measured some indicators of how much actual collaboration happens within some of the web-based collaboration tools that are freely available, such as Swivel and IBM Many Eyes. While limited, the collection of such publicly available data can support basic statistical analysis (Ochoa & Duval, 2008). Our data suggests that some collaboration occurs.
- Finally, we speculate on some of the problems associated with collaborative data analysis that may explain its apparent lack of popularity. For example, data analysis requires a certain level of technical expertise in statistics, mathematics, and information technology that is not necessarily available to everyone. Lack of motivation may also provide an explanation. Other issues may also have harmed people's ability to do collaborative data processing: from human issues such as usability problems or confidentiality, to technical ones such as concurrency, data indexing, or version control.

**REVIEW OF EXISTING TOOLS**

Data processing requires tools specifically built to handle structured data. There are many ways in which people can interact with structured data, and many different tools available to support these various interaction types. Databases are used to store data. Depending on the type of data, it can be manipulated using tools such as spreadsheets or statistical packages. Data can also be transformed and presented in graphical form using visualization tools. In order for groups to collaboratively access structured data, they must use groupware versions of these various tools. We present a few of the available groupware tools here. We also present collaboratories, which are groupware tools aimed specifically at scientists who need to share data.

Recently, many online tools have become available to help people collaboratively analyze data, not only by storing and accessing data, but also by manipulating, analyzing, and visualizing it. Several of these products are web-based enterprise software, which Göldi (2007) has described as a potentially disruptive new model. Specifically, Göldi suggests that web-based tools create new markets: users who have never adopted groupware or collaboratories may use these online tools because they might be simpler or more convenient.

## Database tools

Collaborative data analysis requires a database. In some instances, it may be preferable to give the users direct access to the database where the data is located. When direct access is not required, web database tools can be used. There are many web databases available including Zoho Creator (http://creator.zoho.com/), Dabble DB (http://dabbledb.com/), Quickbase (http://quickbase.intuit.com/), and Caspio Bridge (http://www.caspio.com/). However, none of these products are fully open. These types of tools are more suited for small groups of people collaborating on a project. Google Base (http://base.google.com/) is an open web database, but it fails to offer much with respect to data analysis.

Apache CouchDB, IBM Lotus Notes, and Amazon SimpleDB are software engines for schema-less document-centric databases. The benefit of such an approach is that the data stored in the database does not need to share an agreed-upon schema. In particular, this allows for open disagreement about the meaning of the data stored. However, these engines do not necessarily offer user front-ends to support collaborative data processing. They also do not specifically support end-users.

## Spreadsheet tools

Collaborative spreadsheet tools let several people work on the same spreadsheet either concurrently or in turn. There are several online collaborative spreadsheet tools available: Google Docs Spreadsheets (http://docs.google.com), EditGrid (http://www.editgrid.com/), SecureSheet (http://www.securesheet.com/), SocialCalc (formerly known as wikicalc, http://www.socialcalc.org), and Zoho Sheet (http://sheet.zoho.com). There are also both eXpresso (http://www.expressocorp.com/) and Badblue (http://badblue.com/helpxls.htm) for sharing Microsoft Excel spreadsheets over the Internet. Excel itself includes the functionality to allow people to share a spreadsheet over a LAN.

Academic collaborative online spreadsheet tool projects include TellTable and CoExcel. Both of these transpose a familiar single-user spreadsheet program to a web-based environment (in the first case, the open source Open Office calc, in the latter Microsoft Excel). This offers the interesting advantage that the user is not forced to learn a new tool in order to do their work. TellTable (Adler et al., 2006) is a web-based framework that can turn single user software (including spreadsheets) into collaborative software. TellTable was originally built to provide an audit trail for spreadsheets (Nash et al., 2004). By storing the spreadsheet files on a server (in order to reduce the risk of tampering with data or the audit trail), and making people access the file through the web, the developers were able to transform calc into a groupware spreadsheet program. Only one person at a time can modify a file using TellTable. CoExcel (Sun et al., 2006) is a project to transform Microsoft Excel into a collaborative spreadsheet that can be used over the web. Both TellTable and CoExcel are still in development.

We are not aware of the widespread use of spreadsheet documents as collaborative objects. There may be company-wide read/write spreadsheets but we suspect that the structure forced upon the users of such spreadsheets limits collaboration. For example, inserting a row or a column in a read/write spreadsheet may mistakenly break another user' formula.

## Statistical tools

Like spreadsheet tools, collaborative statistical tools allow people to transform shared data. We have found two online collaborative statistical tools. Statcrunch (http://www.statcrunch.com/)

offers several statistical analyses, including ANOVAs, T tests, regression analyses, and nonparametric statistics. As well, it can display the data using various types of graphics. Covariable (http://www.covariable.com/) also offers various statistical analyses, including T tests, correlations, and linear regressions. Covariable can also display the data in graph form. Both tools allow data and result sharing.

## Visualization tools

Collaborative visualization tools are tools that let people import datasets, create a graph or other type of visual representation of this data, and then share that visualization with other users. There are many such tools available on the web. We describe a few of these here.

DEVise (Livny et al., 1997) is a system that allows users to develop and share visualizations of large datasets. Users can develop their own visual presentations rather than being forced to rely on a collection of pre-existing presentation types (piecharts, etc.). Furthermore, a user can drill all the way down into a visual representation to see an individual data record. Users can share visualizations as well as explore them independently or even concurrently. DEVise has been used for financial, medical, meteorological, biological, and soil sciences datasets. DEVise is meant to support groups working together, but it does not support an open source model. This is also the case for Spotfire (http://spotfire.tibco.com/index.cfm), which is aimed at supporting businesses, and Command Post of the Future (Roth, 2004), which is aimed at the military.

IBM Many Eyes (IBM, 2007) has an open web approach. It is meant specifically to let anyone upload any type of database and share the resulting visualization with everyone on the Web (Viégas et al., 2007, 2008). According to its creators, users' most common activities on the site are "to upload data, construct visualizations, and leave comments on either datasets or visualizations." The authors have included several features specifically to support collaboration, including communication tools such as text comments, annotations, and bookmarks on visualizations. Within the first two months of its life, IBM Many Eyes had gathered over 1,400 users who had uploaded about 2,100 datasets, created 1,700 visualizations and added about 450 comments. These results suggest that people are willing to work together on data analysis.

Swivel (http://www.swivel.com) is somewhat similar to IBM Many Eyes, although the visualization tools it offers are not as powerful (Butler, 2007). Swivel also makes it easy for users to mash datasets together to come up with new and interesting visualizations of their data. Other, similar tools include Data360 (http://www.data360.org) and Trendrr (http://trendrr.com/). Dataplace (http://www.dataplace.org) is aimed more specifically at housing and demographic data from the United States. It lets people quickly create thematic maps by translating data onto maps. Daytum (http://daytum.com) is a social dashboard allowing users to share and visualize data from their daily lives. Finally, Microsoft Research published DataDepot (http://datadepot.msresearch.us), a tool to track and share trend lines generated from data such as precipitation levels or stock prices.

Most of these tools accept standard data formats such as column-separated-values (CSV) files. The results can often be shared as a URI where comments can be added, or new views generated.

## Collaboratories

Collaboratories are collaboration tools aimed specifically at scientists. The term 'collaboratory' first appeared in 1993 (Cerf et al., 1993) and describes a virtual research center, in which scientists from various laboratories across the world can cooperate and share data, resources, and information. While early examples of collaboratories were aimed at giving scientists access to expensive instruments such as particle accelerators (Kouzes, Myers & Wulf, 1996), most of the collaboratories we find today are web-based database repositories (Ma, 2007). Collaboratories are very popular. Already in 2004, the Science of Collaboratories website (http://www.scienceofcollaboratories.org/Resources/colisting.php) noted the existence of over 200 different collaboratories. Recent examples of collaboratories include the National Human Neuroimaging Collaboratory (Keator et al., 2008), the Michigan Clinical Research Collaboratory (Schwenk & Green, 2006) and the Center for Behavioral Neuroscience (Powell & Albers, 2006).

There have been several attempts at building true online collaborative tools for scientists (e.g., Reed, Giles & Catlett, 1997; Avery & Foster, 2000; Reed, 2003; Ma, 2007). These collaboratory projects are usually aimed at very specific groups of research labs who are working together on a project and who require tools to support their collaborative needs.

In this section, we have shown that there are several tools available to do collaborative data analysis but are people willing to use them? The following section explores this question by trying to measure the level of collaboration in some web-based data processing tools.

## EXPERIMENTAL EVALUATION

Golovchinsky et al. (2008, 2009) uses four dimensions to classify computer-supported information seeking collaboration: intent, depth of mediation, concurrency, and location. Because of the way most collaboration is done with the available open access collaborative data analysis tools, we are more concerned here by the intent dimension. Intent is explicit when people get together and work on a specific topic. For example, if coworkers must produce an annual report or research papers, their collaboration has an explicit intent. An example of collaboration with implicit intent is certain recommender systems, in which past search behavior is used to suggest other topics (e.g., Amazon's "people who bought this product also bought").

Collaboration is often multimodal: whereas some users may use a chat tool, others will prefer email (Noël & Robert, 2004; Morris, 2008). This may lead to biases in the trace analysis. For example, Grippa et al. (2006) have shown that relying on email traces alone may overestimate the influence of a core group of individuals who dominate a given mode of communication. Thus, one individual might be a prolific email user whereas his peers use the phone. These biases make it difficult to study groups or projects based only on openly available data.

It might be difficult or impossible to evaluate collaboration between any two given individuals by limited traces, but we may compare one social web site with another and derive some quantifiable information. Minimally, posting content for all to see, especially when others can comment, shows an openness to collaboration. Publicly reacting to existing content is also necessarily a form of collaboration. These types of behavior are a form of global (at the scale of the community) collaboration. In the terms of Golovchinsky et al. (2008, 2009), we have collaboration with implicit intent. A large number of people contributing content indicates that more implicit collaboration is occurring.

Using this method of measuring collaboration on a variety of social web sites, Ochoa and Duval (2008) divided the sites into three categories. In the first category (Amazon Reviews, Digg, FanFiction, and SlideShare), 10% of the users contributed 40% to 60% of the content, in the second category (Furl, LibraryThing, and Revver), 10% of the users contributed 60% to 80% of the content, and in the last category, most of the content was contributed by a few users. The first category of tools are built more collaboratively than the other two categories, in the sense that more users contributed to the content.

It might be surprising that having 10% of the users contribute 50% of the content means that these tools foster a (relatively) large amount of collaboration. As a basis for comparison, Lotka's law states that the number of authors making $n$ scientific contributions is about $1/n^a$ of those making one contribution, where $a$ is about two (Lotka, 1926; Chung & Cox, 1990). One consequence of Lotka's law is that given a large enough set of scientists, almost all contributions (say 90%) will be due to the top 10% most prolific authors.

In which of these three categories do open source database projects fit? We began by looking at OpenWetWare. OpenWetWare is a collection of open science notebooks. While not a data processing site per se, we were interested in this site because it still contains a lot of data regarding biological entities and documentation about data processing techniques. We recovered the last 5000 edits from the site (the data was collected on May 16, 2008). We found that 10% of the users contributed 50% of the changes. Hence, by the Ochoa-Duval categorization, there is a significant amount of collaboration. This suggests that scientists are interested in sharing information with others.

Turning to the open data processing sites, we selected three candidates: IBM Many Eyes, Swivel, and StatCrunch. There are no other publicly available sites similar to these three to our knowledge. We were able to retrieve contributions by different users from all three sites. All operate in a similar manner except that StatCrunch is not free: users must pay a small subscription fee to upload new data and do analyses. However, no subscription is required to view work uploaded to the StatCrunch site. Both StatCrunch and IBM Many Eyes require all contributors to have an account. While Swivel requires users to be registered to upload data, anyone can contribute a new plot.

We captured all data by screen scraping: HTML pages were saved to disk and parsed using regular expressions. To ensure that our scripts did not adversely affect these web sites, we limited our queries to one per second, and we chose to retrieve no more than 1,500 pages from each site. Table 1 presents the size of the datasets. Except for Swivel, our datasets cover thousands of users and tens of thousands of items. Ochoa and Duval (2008) had between 2,300 and 82,000 users per dataset. Both our IBM Many Eyes and StatCrunch datasets are similar in size to their samples.

Swivel offers two ways to navigate through recent plot contributions: by dates and by views. Unfortunately, we found that two thirds of all new plots were from unregistered users. So, for Swivel, we limited our investigation to dataset contributions. We found only 536 users had uploaded datasets.

IBM Many Eyes allows users to browse all recent contributions whether they are new datasets, new plots or new comments. Unfortunately, recent plot contributions are not listed together with the contributing users' ID. Retrieving recent plots might have required loading tens of thousands of web pages, one for each plot. To minimize the impact of our data capture, we limited our investigations to datasets and comments. Of the three websites we investigated, IBM Many Eyes has the largest database, with over 37,000 datasets uploaded by over 8,800 users. Only about 7% of all users posted comments.

StatCrunch allows users to navigate through recent datasets, results, and publicly published reports. Whereas the datasets are raw data, results are the output of some data processing, and the reports are aggregated results. Hence, there is a hierarchy of complexity on StatCrunch, from data to individual results to reports. More users provided datasets than reports. Also, there are far fewer reports than results or datasets. We examined information for datasets, results, and reports from StatCrunch.

*Table 1. Descriptive Statistics of the collected data*

|  |  | Number of items | Number of users |
|---|---|---|---|
| Swivel | Datasets | 1,200 | 536 |
| IBM Many Eyes | Datasets | 37,672 | 8,843 |
|  | Comments | 1,223 | 602 |
| StatCrunch | Datasets | 9,351 | 1,833 |
|  | Results | 11,403 | 1,190 |
|  | Reports | 1,928 | 358 |

For each site, and each chosen type of contribution, we took the 10% most prolific users and counted their contributions. Table 2 presents the results of our analysis. For example, on IBM Many Eyes, the top 10% of users uploaded nearly 19,000 datasets. We did not take into account the volume of each contribution (such as the number of bytes uploaded).

All contributions correspond to the first Ochoa-Duval category: the top 10% of the users contribute 40% to 60% of the content. Hence, at least a moderate amount of collaboration is occurring. The reports in StatCrunch are the only exception. They have an almost flat distribution: the top 10% of all users contribute only 19% of all reports. However, only 358 users published a report compared to 1,833 users who produced at least a dataset (a ratio of 1 to 5).

*Table 2. Percentage of contributions by top 10% most prolific users*

|  |  | Contributions by top 10% of users |
|---|---|---|
| Swivel | Datasets | 48% |
| IBM Many Eyes | Datasets | 49% |
|  | Comments | 55% |
| StatCrunch | Datasets | 57% |
|  | Results | 40% |
|  | Reports | 19% |

In summary, data analysis sites such as Swivel, IBM Many Eyes, and StatCrunch are collaborative, in the sense that the most prolific individuals contribute only about half of the content. They are comparable with popular sites such as Amazon Reviews, Digg, FanFiction, and SlideShare.

We probably underestimate the total number of users of these websites. As we have already mentioned, people can use the systems without leaving a public trace, making them uncounted collaborators. Since these websites let people create private groups, there may be local (at the project-level) collaboration that we are not measuring.

## INGREDIENTS FOR A SUCCESSFUL COLLABORATIVE DATA ANALYSIS

We have shown above that people are willing to collaborate on databases, at least on a global community level, and that there exist plenty of groupware tools that could be used to support this collaboration. Yet, our results suggest that contribution may not be as widespread as it could be. For example, while over 8,000 users created graphs on IBM Many Eyes, only about 600 users contributed comments, and 66 of these users contributed 55% of the comments.

Even when people use collaborative tools to work on a project, data analysis may remain a single-user task, at least in our experience. Nor is there any large-scale equivalent to Wikipedia for structured data. Clearly there are stumbling blocks that are limiting the popularity of collaborative data analysis tools. In this section, we examine some of the issues that designers of these tools need to consider if they wish to promote successful collaboration on structured data.

### Sharing the data

The several large repositories listed above show that there is much data being shared. However, data sharing is not always accepted as a requirement, even in science. For example, Reidpath and Allotey (2001) asked 29 corresponding authors of research articles which appeared in the British Medical Journal to share their data. Only one author actually shared his data in this survey. In the case of businesses, people may be wary of sharing data with their competitors. Many factors may explain why data sharing is not forthcoming: documenting and packaging data for others requires some work; there can be confidentiality and security issues; there is the fear of ridicule if others find errors in one's work; there may be a competitive edge by having data others do not; and so on. However, there is evidence that sharing detailed research data can be beneficial to authors by increasing their citation rate (Piwowar et al., 2007).

Certainly data sharing, when it occurs, may lead to collaborative efforts in the sense that several distinct teams may successively work on overlapping datasets. However, the processing itself may still be done independently (Shah, 2008).

### Task specialization

Local or internal collaboration occurs within closed teams. It is often customary within teams to delegate specific tasks to specific individuals. While some members may handle data processing, others will write text, contribute ideas, or manage the team. If the team is small and relatively stable, there may not be any reason to share the technical task of analyzing the data. Indeed, specialization may be more effective. Among prolific book writers, dividing up the task into chapters is a common strategy (Hartley & Branthwaite, 1989). Posner and Baecker (1993) provide several reasons to specialize the writing tasks including: access to technology and software, social status, familiarity with the requirements, and uniformity of the final product.

### Credit

External or global collaboration occurs more openly. Examples include OpenWetWare where scientists share open notebooks with the world. We believe that external collaboration over data processing is still an outlier in science. According to Hannay, the biggest barrier to collaboration in science is the credit problem (Waldrop, 2008): individuals need to feel certain that their work will not be scooped by others.

## Access

In collaborative writing, Posner and Baecker (1993) have found that access to technology and software sometimes determines the division of labor. Tutt et al. (2007) have presented a generalized concept - local action - wherein some actions can only occur in some locations. With web-based applications we expect this issue to be less significant.

Data is often confidential, making publication and sharing more difficult. This includes personal information and strategic business data. For many users, data privacy remains a necessity and lack of privacy may constitute a barrier to sharing (Descy, 2007). However, a lot of the data being collected is now available freely.

## Expertise

Data analysis sometimes requires a technical expertise in statistics, mathematics, and information technology that few people possess. In addition, learning to use spreadsheet or statistical tools requires time and energy that many people may not have. If data analysis can be delegated to a single person who is already familiar with these tools, then the group will waste less time during this phase. While this argument may hold in a closed social network, like a research group, it is less likely to be true for very large groups, where many people have sufficient expertise for data processing. Wikipedia has shown that large-scale collaboration is not only possible on difficult problems, such as crafting a highly technical article, but that it can also be very fruitful.

## Concurrency and asynchronicity

Version control in situations where changes can be reverted or done in parallel remains challenging (Adler et al. 2006). Keeping track of changes can be particularly difficult with databases as compared with text. Furthermore, because the integrity of data can be a very important issue, an audit trail that allows people to track all changes is an essential tool for collaborative databases.

We can distinguish two types of data: collected or raw data, and derived data. Assuming collaborators agree on the meaning of the collected data, they will build upon it by applying various operations. These operations generate views (derived data), which may come to depend on each other. For example, a user may take the raw data, prune some of the outliers (view 1) and then compute statistical measures on the outlier-free data (view 2). If these views are defined only through algorithms, keywords, or formulae it may become difficult for others to follow the flow. Wattenberg (2005) has underlined the importance of letting people access past data states when sharing visualizations. Therefore any system that lets people work on data needs to not only make people aware of the historical modifications and computations made to the data, but also give them easy access to past states and the ability to work on these states (letting people do 'what if' scenarios with ease). Moreover, since interpretation errors are likely to be common, a feedback loop is needed: one should share interpretations in such a way that they can be corrected by others.

As in text processing, users may need to work jointly on the same object. If the tool permits concurrent editing, this can result in conflict (e.g., if people try to modify the same data at the same time). Several methods solve such potential conflicts during synchronous editing (Mitchell, 1996). However, because text tends to convey meaning better than numbers or figures, conflicts of intent may be more difficult to detect and resolve in databases than in text documents. Even if people are only permitted to work on the same data asynchronously, such conflicts can still arise due to the interdependence of data points.

## Usability

In large datasets, navigation is difficult and retrieval of data may also prove difficult without proper indexing. Performance issues may plague users who have to work with these very large datasets. The volume of data can make sharing more difficult due to bandwidth limitations. Even though bandwidth and storage are increasing, so is the size of the datasets, eliminating any possible gains.

Reorganizing data on the fly is relatively easy using a wiki, since you can simply copy and paste the data. In a collaborative spreadsheet, the data tends to be more rigidly organized (more structured) making on-the-fly reorganization difficult. In addition, users may be unfamiliar with the existing tools. Exchanging reports prepared by a single individual, and commenting using the familiar email interface is often an attractive proposition.

Preparing a convenient output can be difficult since most data processing tasks are not easily exported in a meaningful format to users, contrary to what is possible with collaborative editing tools such as wikis. One partial solution to that problem is to give people the ability to create and export visualizations, which are outputs that are easy to understand.

## Flexible semantics

Traditional data sharing and data integration approaches require a globally consistent data instance (Taylor & Ives, 2006). Current Business Intelligence techniques tend to define schemas and semantics in a centralized manner (Aouiche et al., 2008). However, people often disagree on the semantics of the collected data. This may be especially challenging in cases where the data was collected by remote groups. Even among the group members who collected the data, there may be disagreement on the meaning of the numbers. Inside businesses, there are commonly disagreements on the exact meaning of simple terms like revenue or profit. Improper or ambiguous documentation may prove problematic.

Meanwhile, schemes for large-scale data sharing have generally failed because database approaches tend to impose strict global constraints: a single global schema, a globally consistent data instance, and central administration (Green et al. 2007). Spotfire, Business Objects, and QlikTech are among the companies providing tools to enable the average user to contribute by relaxing global constraints (Havenstein, 2003).

## Motivation

Given access to the right tools, people are willing to engage in social data analysis. For example, Wattenberg (2005) found that people were challenging each other to find trends concerning baby names while using NameVoyager. NameVoyager is a web-based visualization tool (http://www.babynamewizard.com/voyager) that displays the popularity of baby names in the U.S. over time. It covers over 100 years of data. People can drill down to a subset of names by typing in letters; this will display all the names starting with those letters as the letters are typed in. According to Wattenberg, people were building on others' findings, making this an example of group data mining. Wattenberg suggests that if a collaborative tool is to encourage people to share their discoveries it needs to let them easily re-create or access data states previously created by others. The popularity of sites such as IBM Many Eyes and Swivel clearly shows that people are willing to play around with complex data and share their results with the world. Another example of such a site is Mycrocosm, a website published by MIT researchers where users can share the "minutiae of daily life" as simple statistical graphs (http://mycro.media.mit.edu/).

Data analysis may not be perceived as interesting but the ubiquity of statistics in our daily life combined with appealing visualization techniques should alleviate any prejudice. Moreover, programming is also a rather technical task and open source software has shown that open collaboration is a working model for software creation. Prolific contributors to Wikipedia are motivated by increased credibility within the community (Forte & Bruckman, 2005) as well as by altruistic goals such as contributing to the greater good (Wagner & Prasarnphanich, 2007). Participation in open source software is similarly motivated (Wu et al., 2007) by altruistic goals and also by possible career advancement. These same incentives are likely to be present for collaborative data processing. To ensure that individuals get adequate credit for their work, tools should track the authors of various contributions and make this information accessible to users.

## Conclusion

Ioannidis (2005) claimed that most research results are wrong because most datasets are too small or the investigation is biased. He observed that researchers commonly select the most positive results and discard the negative results prior to publication. These biases may be exacerbated if there are commercial imperatives underlying the work. There is no reason to believe that the same biases are not present in Business Intelligence. If a more diverse set of people could analyze the same data, it seems likely that biases in the analysis would be less frequent. Collaborative data processing tools could help support these multiple analyses. Would open collaborative data processing actually increase the reliability of published results? Given large data repositories, should we set up collaborative data processing initiatives? Should companies rely on a wider range of employees to process their data? While the reliability of open encyclopedias such as Wikipedia is often reported to be good, noise and missing information are concerns (Clauson et al., 2008). However, since open collaborative data processing can contribute to making research reproducible, it might make it easier to detect some types of fraudulent behavior (Laine et al. 2007). We believe that other benefits to open collaborative data analysis are likely to exist, but it remains to identify under which circumstances this type of data analysis is most likely to be beneficial. Ultimately, we need better experimental evaluation of collaborative data analysis, including longitudinal studies.

A distinguishing factor between collaboration over a wiki or software, and collaborative data analysis might be the barrier to entry. Many scientists and business analysts may have a vested interest in controlling not only access to their data, but also the analytical process. A cultural change wherein access to the raw data and to the processing steps would be an essential part of any report or scientific communication, would likely be disruptive in the same way Wikipedia has been disruptive within the encyclopedia industry and open source software disruptive within the software industry. Would collaborative data processing lead us to a new form of highly collaborative science?

There are many good collaborative software packages, and a substantial amount of good collaboration. None of the limitations to collaborative data processing that we have identified are entirely unsolved. Nevertheless, the sophistication of the open data processing tools could be greater. Tools such as IBM Many Eyes or Swivel are appropriate for generating graphics, but they cannot process data. As well, while they permit implicit collaboration, it is not clear how well they support explicit types of collaboration. Shareable spreadsheets are familiar to users, but were designed for single users. Based on our survey and on earlier work (Adler et al., 2006; Aouiche et al., 2008), here are some features that we recommend:

1. Due to the need for flexible semantics, data should be presented in an unstructured manner. We should encourage loose couplings in how data is presented to the users: large structured tables or graphs are probably not appropriate. We expect that it is difficult to scale a spreadsheet to dozens of simultaneous users.
2. In the spirit of tools such as IBM Many Eyes and Swivel, the result of any editing should be immediately shareable in an output easily understood by human readers. Ideally, any result should be shareable as a single URI.
3. Users should be able to go back to past data states and branch out from there to new data analyses.
4. Visualizations can simplify the presentation of complex data. Therefore, some sort of visualization tool should be included.
5. Some type of peer review process may be required to control data quality. Wikipedia offers a working example of group quality control that depends on people, not on complex tools.
6. Any change should be clearly credited to its author.
7. To alleviate the local action problem, we should use open standards to enter data and to publish data analysis.

## ACKNOWLEDGEMENTS

The authors wish to thank Sarah Dumoulin and two anonymous reviewers for their suggestions and comments. The second author is supported by NSERC grant 261437.

## REFERENCES

Adler, A., Nash, J. C., & Noël, S. (2006). Evaluating and implementing a collaborative office document system. *Interacting with Computers*, 18 (4), 665-682.

Aouiche, K., Lemire, D., & Godin, R. (2008). Collaborative OLAP with tag clouds: Web 2.0 OLAP formalism and experimental evaluation. *Proceedings of the 4th International Conference on Web Information Systems and Technologies (WEBIST 2008)*. Funchal, Madeira, Portugal, 5-12.

Avery, P., & Foster, I. (2000). The GriPhyN Project: Towards petascale virtual-data grids. GryPhyn Report 2000-1. Retrieved February 9, 2009, from http://www.griphyn.org/documents/document_server/uploaded_documents/doc--501--proposal_all.doc

Butler, D. (2005). Science in the web age: Joint efforts. *Nature*, *438*, 548-549.

Butler, D. (2007). Data sharing: The next generation. *Nature*, *446*, 10-11.

Canadian Institutes of Health Research (2007). Access to research outputs. Retrieved February 9, 2009, from http://www.cihr-irsc.gc.ca/e/34846.html

Cerf, V. G., Cameron, A., Lederberg, J., Russel, C., Schatz, B., Shames, P. et al. (1993). *National Collaboratories: Applying Information Technologies for Scientific Research*, Washington, D.C.: National Academy Press.

Chung, K. H., & Cox, R. A. K. (1990). Patterns of productivity in the finance literature: A study of the bibliometric distributions. *Journal of Finance*, *45* (1), 301–309.

Clauson, K.A., Polen, H.H., Boulos, M.N.K., & Dzenowagis, J.H. (2008). Scope, completeness, and accuracy of drug information in Wikipedia. *The Annals of Pharmacotherapy*, *42* (12), 1814.

Descy, D. E. (2007). Browser-based online applications: Something for everyone! *TechTrends: Linking Research and Practice to Improve Learning*, *51* (2), 3-5.

Dignan, L. (2007). Salesforce.com rolls out customer data sharing; eyes 1 million subscribers. *ZDNet*. Retrieved February 9, 2009, from http://blogs.zdnet.com/BTL/?p=7239

Findley, L. G. R. & Inge, L. G. J. (2005). North American defence and security in the aftermath of 9/11. *Canadian Military Journal*, *6* (1), 9-16.

Forte, A., & Bruckman, A. (2005). Why do people write for Wikipedia? Incentives to contribute to open-content publishing. Paper presented at the *GROUP* workshop "Sustaining community: The role and design of incentive mechanisms in online systems", Sanibel Island, FL. Retrieved 9 February, 2009, from http://www.cc.gatech.edu/~aforte/ForteBruckmanWhyPeopleWrite.pdf

Gartner Inc. (2007, January 30). Business intelligence market will grow 10 percent in EMEA in 2007 according to Gartner [Press release]. Retrieved February 9, 2009, from http://www.gartner.com/it/page.jsp?id=500680.

Göldi, A. (2007). The Emerging Market for Web-based Enterprise Software, Unpublished master's thesis: Massachusetts Institute of Technology, Boston, Mass.

Golovchinsky, G., Pickens, J. & Back, M. (2008). A taxonomy of collaboration in online information seeking. *11th International Workshop on Collaborative Information Retrieval, JCDL 2008*, June 20, 2008.

Golovchinsky, G., Qvarfordt, P. & Pickens, J. (2009). Collaborative information seeking. *Computer*, March, 47- 51.

Green, T. J., Karvounarakis, G., Taylor, N.E., Biton, O., Ives, Z. G., & Tannen, V., (2007). ORCHESTRA: Facilitating collaborative data sharing. *Proceedings of the ACM International Conference on Management of Data (SIGMOD'07)*, Beijing, China, 1131-1133.

Grippa, F. , Zilli, A., Laubacher, R., & Gloor, P. (2006). E-mail may not reflect the social network. *Proceedings of the 2006 International Sunbelt Social Network Conference*, Vancouver, BC, Canada.

Hartley, J., & Branthwaite, A. (1989). The psychologist as wordsmith: A questionnaire study of the writing strategies of productive British psychologists. *Higher Education*, *18* (4), 423-452.

Havenstein, H. (2003). BI vendors seek to tap end-user power. *InfoWorld*, *25* (22).

IBM, Inc. (2007). Many Eyes. Retrieved February 9, 2009, from http://manyeyes.alphaworks.ibm.com/manyeyes/

Ioannidis, J. P. (2005). Why most published research findings are false. *PLoS Medicine*, *2* (8), e124.

Jones, C. (2007). Intelligence reform: The logic of information sharing. *Intelligence & National Security*, *22* (3), 384-401.

Keator, D. B., Grethe, J.S., Marcus, D., Ozyurt, B., Gadde, S., Murphy, S. et al. (2008). A national human neuroimaging collaboratory enabled by the Biomedical Informatics Research Network (BIRN). *IEEE Transactions on Information Technology in Biomedicine*, *12* (2), 162-172.

Kouzes, R. T., Myers, J. D., & Wulf, W. A. (1996). Collaboratories: Doing science on the Internet. *Computer*, *29* (8), 40-46.

Laine, C., Goodman, S. N., Griswold, M. E., & Sox, H. C. (2007). Reproducible research: Moving toward research the public can really trust. *Annals of Internal Medicine*, *146* (6), 450-453.

Livny, M., Ramakrishnan, R., Beyer, K., Chen, G., Donjerkovic, D., Lawande, S. et al. (1997). DEVise: Integrated querying and visual exploration of large datasets. *Proceedings of the ACM International Conference on Management of Data (SIGMOD'97)*, 301-312.

Lotka, A. J. (1926). The frequency distribution of scientific productivity. *Journal of the Washington Academy of Science*s, *16* (12), 317–324.

Lund University Libraries (2003). Directory of open access journals. Retrieved February 9, 2009, from http://www.doaj.org/

Ma, K.-L. (2007). Creating a collaborative space to share data, visualization and knowledge. *ACM SIGGRAPH Computer Graphics Quarterly*, *41* (4).

Mitchell, A. (1996). Communication and Shared Understanding in Collaborative Writing. Unpublished Master's Thesis. Computer Science Department, University of Toronto.

Morris, M. R. (2008). A survey of collaborative web search practices. Proceedings of the ACM Conference on Computer-Human Interaction (CHI'08), 1657-1660.

Noël, S., & Robert, J.-M. (2004). Empirical study on collaborative writing: What do co-authors do, use and like? *Computer Supported Cooperative Work*, *13* (1), 63-89.

Nash, J., Adler, A., & Smith, N. (2004). TellTable spreadsheet audit: From technical possibility to operating prototype. *Proceedings 2004 Conference European Spreadsheet Interest Group*, 45-56.

Niu, J. (2006). Incentive study for research data sharing. Retrieved February 9, 2009, from http://icd.si.umich.edu/twiki/pub/ICD/LabGroup/fieldpaper_6_25.pdf

Ochoa, X., & Duval, E. (2008). Quantitative analysis of user-generated content on the Web. *Web Science Workshop WebEvolve*.

Okada, K. I., Maeda, F., Ichikawaa, Y., & Matsushita, Y. (1994). Multiparty videoconferencing at virtual social distance: MAJIC design. *Proceedings 1994 ACM conference on Computer supported cooperative work*, 385-393.

Pedersen, E. R., McCall, K., Moran, T. P., & Halasz, F. G. (1993). Tivoli: An electronic whiteboard for informal workgroup meetings. *Proceedings SIGCHI conference on Human factors in computing systems*, 391-398.

Piwowar, H. A., Day, R. S., & Fridsma, D. B. (2007). Sharing Detailed Research Data Is Associated with Increased Citation Rate. *PLoS ONE*, 2(3).

Posner, I., & Baecker, R. M. (1993). How people write together. In R.M. Baecker (Ed.), *Readings in Groupware and Computer-Supported Cooperative Work: Assisting Human-Human Collaboration*. San Mateo, CA: Morgan Kaufman, pp. 239-250.

Powell, K. R. & Albers, H. E. (2006). Center for Behavioral Neuroscience: A prototype multi-institutional collaborative research center. *Journal of Biomedical Discovery and Collaboration*, *1* (1), 9.

Reed, D. A. (2003). Grids, the TeraGrid, and Beyond. *Computer*, *36* (1), 62-68.

Reed, D. A., Giles, R. C., & Catlett, C. E. (1997). Distributed data and immersive collaboration. *Communications of the ACM*, *40* (11), 38-48.

Reidpath, D. D., & Allotey, P. A. (2001). Data Sharing in Medical Research: An Empirical Investigation. *Bioethics*, *15* (2), 125-134.

Roth, S. (2004). Capstone Address: Visualization as a medium for capturing and sharing thoughts. *Proceedings IEEE InfoVis 2004*, xiii.

Schwenk, T. L., & Green, L. A. (2006). The Michigan Clinical Research Collaboratory: Following the NIH roadmap to the community. *Annals of Family Medicine*, *4* (1), 49-54.

Shah, C. (2008). Toward Collaborative Information Seeking (CIS). *Proceedings of the Collaborative Exploratory Search Workshop*.

Stein, L. D., Mungall, C., Shu, S.Q., Caudy, M., Mangone, M., Day, A. et al. (2002). The Generic Genome Browser: A building block for a model organism system database. *Genome Research*, *12*, 1599-1610.

Sun, C., Xia, S., Sun, D., Chen, D., Shen, H. & Cai, W. (2006). Transparent adaptation of single-user applications for multi-user real-time collaboration. *ACM Transactions on Computer-Human Interaction*, *13* (4), 531-582.

Swivel, Inc. (2007). Swivel. Retrieved February 9, 2009, from http://www.swivel.com

Taylor, N. E., & Ives, Z. G. (2006). Reconciling while tolerating disagreement in collaborative data sharing. *Proceedings of the ACM Conference on Management of Data (SIGMOD '06)*, 13-24.

Tutt, D., Hindmarsh, J., & Fraser, M. (2007). The distributed work of local action: Interaction amongst virtually collocated research teams. *Proceedings of the European Conference on Computer-Supported Cooperative Work (ECSCW 2007)*, 199-218.

Viégas, F. B., Wattenberg, M., van Ham, F., Kriss, J., & McKeon, M. (2007). Many Eyes: A site for visualization at internet scale. *Proceedings Infovis 2007*, 1121-1128.

Viégas, F. B., Wattenberg, M., McKeon, M., van Ham, F. , & Kriss, J. (2008). Harry Potter and the meat-filled freezer: A case study of spontaneous usage of visualization tools. *Proceedings HICSS 2008*, 159.

Wagner, C., & Prasarnphanich, P. (2007). Innovating collaborative content creation: The role of altruism and wiki technology. *Proceedings HICSS 2007*, 40 (1), 278.

Waldrop, M. M. (2008). Science 2.0 - Is Open Access Science the Future? *Scientific American*, Retrieved February 9, 2009, from http://www.sciam.com/article.cfm?id=science-2-point-0

Wattenberg, M. (2005). Baby names, visualization, and social data analysis. Proceedings Infovis 2005, 1-7.

Wilbanks, J. & Boyle, J. (2006), Introduction to Science Commons. Retrieved February 23, 2009, from http://sciencecommons.org/wp-content/uploads/ScienceCommons_Concept_Paper.pdf

Wu, C. G., Gerlach, J. H., & Young, C. E. (2007). An empirical analysis of open source software developers' motivations and continuance intentions. *Information & Management*, *44* (3), 253-262.

## ADDITIONAL READING

Adler, A., & Nash, J. C. (2004). Knowing what was done: uses of a spreadsheet log file. *Spreadsheets in Education*, *1* (2), 118-130.

Arita, M., & Suwa, K. (2008). Search extension transforms wiki into a relational system: A case for flavonoid metabolite database. *BioData Mining*, *1* (7).

Bentley, R., Horstmann, T., & Trevor, J. (1997). The World Wide Web as enabling technology for CSCW: The case of BSCW. *Computer Supported Cooperative Work*, *6* (2-3), 111-134.

Cederqvist, P. (2002). *Version management with CVS*. Bristol UK: Network Theory.

Cole, P., & Nast-Cole, J. (1992). A primer on group dynamics for groupware developers. In D. Marca and G. Bock (Eds.), *Groupware: Software for Computer-Supported Cooperative Work* (pp. 44-57). Los Alamitos, CA: IEEE Computer Society Press.

Dillon, A., & Maynard, S. (1995). 'Don't forget to put the cat out' - Or why collaborative authoring software and everyday writing pass one another by! *The New Review of Hypermedia and Multimedia*, *1*, 135-153.

Dix, A. (1997). Challenges for Cooperative Work on the Web: An Analytical Approach. *Computer Supported Cooperative Work*, *6* (2), 135-156.

Elbashir, M. Z., Collier, P. A., & Lee, S.-F. (2008). Measuring the effects of business intelligence systems: The relationship between business process and organizational performance. *International Journal of Accounting Information Systems*, *9* (3), 135-153.

Grigori, D., Casati, F., Castellanos, M., Dayal, U., Sayal, M., & Shan, M.C. (2004). Business Process Intelligence. *Computers in Industry*, *53* (3), 321-343.

Grudin, J. (1992). Why CSCW applications fail: Problems in the design and evaluation of organisational interfaces. In D. Marca and G. Bock (Eds), *Groupware: Software for Computer-Supported Cooperative Work* (pp. 552-560). Los Alamitos, CA: IEEE Computer Society Press.

Grudin, J. (1994). Groupware and social dynamics: Eight challenges for developers. *Communications of the ACM, 37* (1), 92-105.

Grudin, J., & Palen, L. (1995). Why Groupware Succeeds: Discretion or Mandate? *Proceedings of the European Conference on Computer-Supported Cooperative Work (ECSCW'95)*, 263-278.

Hodis, E., Prilusky, J., Martz, E., Silman, I., Moult, J., & Sussman, J.L. (2008). Proteopedia - A scientific 'wiki' bridging the rift between three-dimensional strucutre and function of biomacromolecules. *Genome Biology*, *9*, R121.

Hoffmann, R. (2008). A wiki for the life sciences where authorship matters. *Nature Genetics*, *40*, 1047-1051.

Kaser, O. & Lemire, D. (2007). Tag-Cloud Drawing: Algorithms for Cloud Visualization. Paper presented at the WWW2007 Workshop: Tagging and Metadata for Social Information Organization.

Leuf, B., & Cunningham, W. (2001). *The Wiki Way: Quick Collaboration*. Addison-Wesley.

Noël, S., & Robert, J.-M. (2003). How the Web is used to support collaborative writing. *Behaviour & Information Technology*, *22* (4), 245-262.

Powel, S. G., Baker, K. R., & Lawson, B. (2007). An auditing protocol for spreadsheet models. *Information & Management*, *45* (5), 312-320.

Powel, S. G., Baker, K. R. and Lawson, B. (2008). A critical review of the literature on spreadsheet errors. *Decision Support Systems*, *46* (1), 128-138.

Rao, V. S., McLeod, P. L., & Beard, K. M. (1996). Adoption patterns of lowstructure groupware: Experiences with collaborative writing software. *Proceedings HICSS'96*, 41-50.

Rivadeneira, A. W., Gruen, D. M., Muller, M. J., & Millen, D. R. (2007). Getting our head in the clouds: toward evaluation studies of tagclouds. *Proceedings of the ACM Conference on Computer-Human Interaction (CHI'07)*, 995–998.

Russell, T. (2006). Cloudalicious: folksonomy over time. *Proceedings of the Joint Conference on Digital Libraries (JCDL'06)*, 364–364.

Wang, X. (2008). miRDB: A microRNA target prediction and functional database with a wiki interface. *RNA*, *14* (6), 1012-1017.

Wattenberg, M., & Kriss, J. (2006). Designing for social data analysis. *IEEE Transactions on Visualization and Computer Graphics*, *12* (4), 549–557.

Wu, P., Sismanis, Y., & Reinwald, B. (2007). Towards keyword-driven analytical processing. *Proceedings of the ACM Conference on Management of Data (SIGMOD'07)*, 617–628.